\documentstyle[12pt]{article}
\begin{document}
\title{The Three -- Dimensional $Z_{2}$ Electrodynamics and the 
Two -- Dimensional Ising Model}

\author{Yury M. Zinoviev\thanks{This work is supported in part by the
Russian Foundation for Basic Research (Grant No. 00 -- 01 -- 00083)}\\
Steklov Mathematical Institute, \\ Gubkin St. 8, Moscow 117966, GSP - 1,
Russia \\ e -- mail: zinoviev@mi.ras.ru}

\date{}

\maketitle

\vskip 1cm

\noindent {\bf Abstract.} The correlation functions are calculated for the
three -- dimensional ${\bf Z}_{2}$ electrodynamics for the particular values
of the interaction energies and for the free boundary conditions.

\vskip 1cm

\section{Introduction}

\noindent Onsager \cite{1} invented the formula for the partition function
of the two -- dimensional Ising model. Another method for the calculation
of the partition function was proposed by Kac and Ward \cite{2}. They
considered two formulae simultaneously: the determinant of the special
matrix $I + T$ ($I$ is the identity matrix) is proportional to the partition 
function of the two -- dimensional Ising model and it is proportional to the 
square of the partition function. For the proof of the first formula they 
used a topological statement. Sherman \cite{3}, \cite{4} constructed a
counter -- example for this statement and gave some arguments for the
equality
\begin{equation}
\label{1.1}
Z^{2} = C(\beta )\det (I + T)
\end{equation}
where $Z$ is the partition function of the two -- dimensional Ising model
with the free boundary conditions and $C(\beta )$ is the positive function
of the inverse temperature $\beta $. In the paper \cite{5} the following
formula 
\begin{equation}
\label{1.2}
Z^{2} = C(\beta )\det (I - T)
\end{equation}
is proved for an arbitrary lattice lying on the plane. For the rectangular
lattice the expression (\ref{1.2}) is independent of the sign of the matrix
$T$. For an arbitrary lattice the formula (\ref{1.1}) is wrong. In the paper
\cite{6} the free energy and the correlation functions of the two --
dimensional Ising model with the free boundary conditions are calculated
by using the formula (\ref{1.2}).

In this paper we study the connection between the three -- dimensional
${\bf Z}_{2}$ electrodynamics and the two -- dimensional Ising model.
We consider a finite three -- dimensional cubic lattice. The vertices of
this lattice are given by the vectors ${\bf p} = (p_{1},p_{2},p_{3})$
where the numbers $p_{i}$, $i = 1,2,3$, are natural. The non -- oriented
edge of this lattice is given by the pair $\{ {\bf p},{\bf e}\} $ where
${\bf p}$ is the edge initial vertex and the unit vector ${\bf e}$ is the
edge direction. The unit vector ${\bf e}$ is one of six vectors:
$(\pm 1,0,0)$, $(0,\pm 1,0)$, $(0,0,\pm 1)$. Since the edge is non -- 
oriented, then $\{ {\bf p} + {\bf e}, - {\bf e}\} = \{ {\bf p},{\bf e}\} $.
The non -- oriented face of this lattice is given by the triplet
$\{ {\bf p},{\bf e}_{1},{\bf e}_{2}\} $ where ${\bf p}$ is the face initial
vertex and the unit vectors ${\bf e}_{1},{\bf e}_{2}$ are two orthogonal
to each other vectors from six unit vectors $(\pm 1,0,0)$, $(0,\pm 1,0)$, 
$(0,0,\pm 1)$. Since the face is non -- oriented, then 
$\{ {\bf p},{\bf e}_{1},{\bf e}_{2}\} = \{ {\bf p},{\bf e}_{2},{\bf e}_{1}\} =
\{ {\bf p} + {\bf e}_{1}, - {\bf e}_{1},{\bf e}_{2}\} =
\{ {\bf p} + {\bf e}_{2},{\bf e}_{1}, - {\bf e}_{2}\} =
\{ {\bf p} + {\bf e}_{1} + {\bf e}_{2}, - {\bf e}_{1}, - {\bf e}_{2}\}  $.
By ${\bf Z}^{add}_{2}$ we denote the group of modulo $2$ residuals. The
modulo $2$ residuals are mulptiplied by each other and the group
${\bf Z}^{add}_{2}$ is the field. We consider the function 
$A(\{ {\bf p},{\bf e}\} )$ on the non -- oriented edges taking the values
in the group ${\bf Z}^{add}_{2}$. The energy of ${\bf Z}_{2}$
electrodynamics has the following form
\begin{equation}
\label{1.3}
H(A) = - \sum_{\{ p,e_{1},e_{2}\}}
E(\{ {\bf p},{\bf e}_{1},{\bf e}_{2}\})
(- 1)^{A(\{ p,e_{1}\} ) + A(\{ p + e_{1},e_{2}\} ) +
A(\{ p + e_{2},e_{1}\} ) + A(\{ p,e_{2}\} )}
\end{equation}
where the summing runs over all non -- oriented  faces of the lattice.
The number 

\noindent
$E(\{ {\bf p},{\bf e}_{1},{\bf e}_{2}\})$ is called the
interaction energy attached to the non -- oriented face 
$\{ {\bf p},{\bf e}_{1},{\bf e}_{2}\}$. For the infinite interaction
energy the energy (\ref{1.3}) is also infinite. However, it is possible
to consider the infinite values of the interaction energies
$E(\{ {\bf p},{\bf e}_{1},{\bf e}_{2}\})$ for the correlation functions
and for the quantity
\begin{equation}
\label{1.4}
\left( \prod_{\{ p,e_{1},e_{2}\}} 
\cosh \beta E(\{ {\bf p},{\bf e}_{1},{\bf e}_{2}\})\right)^{- 1}
\sum_{A} \exp \{ - \beta H(A)\}.
\end{equation}
(The product in (\ref{1.4}) runs over all non -- oriented faces of the
lattice. The sum in (\ref{1.4}) runs over all ${\bf Z}^{add}_{2}$ valued
functions on the non -- oriented edges of the lattice.) We will show that
for the interaction energies $E(\{ {\bf p},(1,0,0),(0,1,0)\} )$ equaled
$+ \infty $ (or $- \infty $) the correlation functions and the quantity
(\ref{1.4}) for ${\bf Z}_{2}$ electrodynamics are related to the correlation
functions and the partition function of the two -- dimensional Ising model
with the free boundary conditions.

In the second section the definitions of the partition function and the
correlation functions of the three -- dimensional ${\bf Z}_{2}$
electrodynamics with the free boundary conditions are given. These
quantities are calculated for the for the interaction energies
$E(\{ {\bf p},(1,0,0),(0,1,0)\} ) = 0$. The third section is devoted 
to the connection between the quantity (\ref{1.4}) for the interaction
energies $E(\{ {\bf p},(1,0,0),(0,1,0)\} ) = \pm \infty $ and the
partition function of the two -- dimensional Ising model with the free
boundary conditions. In the fourth section we study the connection 
between the correlation functions of the three -- dimensional ${\bf Z}_{2}$
electrodynamics with the interaction energies
$E(\{ {\bf p},(1,0,0),(0,1,0)\} ) = \pm \infty $ and the correlation 
functions of the two -- dimensional Ising model with the free boundary
conditions.

\section{${\bf Z}_{2}$ Electrodynamics}
\setcounter{equation}{0}

\noindent We consider a rectangular lattice formed by the points with
the integral Cartesian coordinates $x = k_{1}$, $y = k_{2}$, $z = k_{3}$,
$M^{\prime}_{i} \leq k_{i} \leq M_{i}$, $i = 1,2,3$, and the corresponding
edges connecting the neighbour vertices. We denote this graph by
$G_{3} = G(M^{\prime}_{1},M^{\prime}_{2},M^{\prime}_{3};M_{1},M_{2},M_{3})$.
The cell complex $P(G_{3})$ is called the set consisting of the cells
(vertices, edges, faces, cubes). A vertex of $P(G_{3})$ is called a cell
of dimension $0$. It is denoted by $s^{0}_{i}$. An edge of $P(G_{3})$ is
called a cell of dimension $1$. It is denoted by $s^{1}_{i}$. A face of
$P(G_{3})$ is called a cell of dimension $2$. It is denoted by $s^{2}_{i}$.
A cube of $P(G_{3})$ is called a cell of dimension $3$. It is denoted by
$s^{3}_{i}$. To every pair of the cells $s^{p}_{i}, s^{p - 1}_{j}$ there
corresponds the number $(s^{p}_{i}:s^{p - 1}_{j}) \in {\bf Z}^{add}_{2}$
(incidence number). If the cell $s^{p - 1}_{j}$ is included into the
boundary of the cell $s^{p}_{i}$, then the incidence number
$(s^{p}_{i}:s^{p - 1}_{j}) = 1$. Otherwise the incidence number
$(s^{p}_{i}:s^{p - 1}_{j}) = 0$. For any pair of the cells 
$s^{p+ 2}_{i},s^{p}_{j}$ the incidence numbers satisfy the condition
\begin{equation}
\label{2.1}
\sum_{m} (s^{p + 2}_{i}:s^{p + 1}_{m})(s^{p + 1}_{m}:s^{p}_{j}) = 0 \,
\hbox{mod} \, 2.
\end{equation}
Indeed, if the vertex $s^{0}_{j}$ is not included into the boundary of
the square $s^{2}_{i}$ , then the condition (\ref{2.1}) is fulfilled. 
If the vertex $s^{0}_{j}$ is included into the boundary of the square
$s^{2}_{i}$, then it is included into the boundaries of the six edges 
$s^{1}_{m}$ two of which are included into the boundary of the square
$s^{2}_{i}$. The condition (\ref{2.1}) is fulfilled again. If the edge
$s^{1}_{j}$ is not included into the boundary of the cube $s^{3}_{i}$,
then the condition (\ref{2.1}) is fulfilled. If the edge $s^{1}_{j}$ is
included into the boundary of the cube $s^{3}_{i}$, then it is included
into the boundaries of the four squares $s^{2}_{m}$ two of which are 
included into the boundary of the cube $s^{3}_{i}$. The condition 
(\ref{2.1}) is fulfilled again.

A cochain $c^{p}$ of the complex $P(G_{3})$ with the coefficients in the
group ${\bf Z}^{add}_{2}$ is a function on the $p$ -- dimensional cells
taking values in the group ${\bf Z}^{add}_{2}$. Usually the cell
orientation is considered and the cochains are antisymmetric functions:
$c^{p}(- s^{p}_{i}) = - c^{p}(s^{p}_{i})$. However, 
$- 1 = 1 \, \hbox{mod} \, 2$ and we can neglect the cell orientation for
the coefficients in the group ${\bf Z}^{add}_{2}$. The cochains form an
Abelian group 
\begin{equation}
\label{2.2}
(c^{p} + c^{\prime p})(s^{p}_{i}) = c^{p}(s^{p}_{i}) + 
c^{\prime p}(s^{p}_{i}) \, \hbox{mod} \, 2.
\end{equation}
It is denoted by $C^{p}(P(G_{3}),{\bf Z}^{add}_{2})$. The mapping
\begin{equation}
\label{2.3}
\partial c^{p}(s^{p - 1}_{i}) = \sum_{j} (s^{p}_{j}:s^{p - 1}_{i})
c^{p}(s^{p}_{j}) \, \hbox{mod} \, 2
\end{equation}
defines the homomorphism of the group $C^{p}(P(G_{3}),{\bf Z}^{add}_{2})$
into the group $C^{p - 1}(P(G_{3}),{\bf Z}^{add}_{2})$. It is called the
boundary operator. The mapping
\begin{equation}
\label{2.4}
\partial^{\star} c^{p}(s^{p + 1}_{i}) = \sum_{j} (s^{p + 1}_{i}:s^{p}_{j})
c^{p}(s^{p}_{j}) \, \hbox{mod} \, 2
\end{equation}
defines the homomorphism of the group $C^{p}(P(G_{3}),{\bf Z}^{add}_{2})$
into the group $C^{p + 1}(P(G_{3}),{\bf Z}^{add}_{2})$. It is called the
coboundary operator. The condition (\ref{2.1}) implies 
$\partial \partial = 0$, $\partial^{\star} \partial^{\star} = 0$. The
kernel $Z_{p}(P(G_{3}),{\bf Z}^{add}_{2})$ of the homomorphism (\ref{2.3})
on the group $C^{p}(P(G_{3}),{\bf Z}^{add}_{2})$ is called the group of
cycles of the complex $P(G_{3})$ with the coefficients in the group
${\bf Z}^{add}_{2}$. The image $B_{p}(P(G_{3}),{\bf Z}^{add}_{2})$ of
the homomorphism (\ref{2.3}) in the group $C^{p}(P(G_{3}),{\bf Z}^{add}_{2})$
is called the group of boundaries of the complex $P(G_{3})$ with the
coefficients in the group ${\bf Z}^{add}_{2}$. Since $\partial \partial = 0$,
the group $B_{p}(P(G_{3}),{\bf Z}^{add}_{2})$ is the subgroup of the 
group $Z_{p}(P(G_{3}),{\bf Z}^{add}_{2})$. Analogously, for the coboundary
operator $\partial^{\star} $ the group of cocyles 
$Z^{p}(P(G_{3}),{\bf Z}^{add}_{2})$ and the group of coboundaries
$B^{p}(P(G_{3}),{\bf Z}^{add}_{2})$ are defined. 

It is possible to introduce the bilinear form on 
$C^{p}(P(G_{3}),{\bf Z}^{add}_{2})$:
\begin{equation}
\label{2.5}
\langle f^{p},g^{p}\rangle = \sum_{i} f^{p}(s^{p}_{i})g^{p}(s^{p}_{i}) 
\, \hbox{mod} \, 2.
\end{equation}
The definitions (\ref{2.3}), (\ref{2.4}) imply
\begin{eqnarray}
\label{2.6}
\langle f^{p},\partial^{\star} g^{p - 1}\rangle =
\langle \partial f^{p},g^{p - 1}\rangle \nonumber \\
\langle f^{p},\partial g^{p + 1}\rangle =
\langle \partial^{\star} f^{p},g^{p + 1}\rangle .
\end{eqnarray}

Let a cochain $ A^{1} \in C^{1}(P(G_{3}),{\bf Z}^{add}_{2})$. Let the
energy be expressed in the form
\begin{equation}
\label{2.7}
H^{\prime}(\partial^{\star} A^{1}) = \sum_{s^{2}_{i} \in P(G_{3})}
h_{i}(\partial^{\star} A^{1}(s^{2}_{i}))
\end{equation}
where $h_{i}(\epsilon )$ is an arbitrary function on the group
${\bf Z}^{add}_{2}$:
\begin{equation}
\label{2.8}
h_{i}(\epsilon ) = D_{i} - E_{i}(- 1)^{\epsilon}
\end{equation}
and the constants
$$
D_{i} = \frac{1}{2} (h_{i}(1) + h_{i}(0)),
$$
$$
E_{i} = \frac{1}{2} (h_{i}(1) - h_{i}(0)).
$$
The substitution of the equality (\ref{2.8}) into the equality (\ref{2.7})
gives
\begin{equation}
\label{2.9}
H^{\prime}(\partial^{\star} A^{1}) = \sum_{s^{2}_{i} \in P(G_{3})} D_{i}
+ H(\partial^{\star} A^{1})
\end{equation}
where the function
\begin{equation}
\label{2.10}
H(\partial^{\star} A^{1}) = - \sum_{s^{2}_{i} \in P(G_{3})}
E_{i}(- 1)^{\partial^{\star} A^{1}(s^{2}_{1})}
\end{equation}
is called the energy of ${\bf Z}_{2}$ electrodynamics. The number
$E_{i} = E(s^{2}_{i})$ is the interaction energy attached to the face 
$s^{2}_{i}$ .

The function
\begin{equation}
\label{2.11}
Z_{G_{3}} = \sum_{A^{1} \in C^{1}(P(G_{3}),Z^{add}_{2})}
\exp \{ - \beta H(\partial^{\star} A^{1})\}
\end{equation}
is called the partition function of ${\bf Z}_{2}$ electrodynamics.

Let the cochain $\chi^{1} \in C^{1}(P(G_{3}),{\bf Z}^{add}_{2})$ take the
value $1$ at the edges $s^{1}_{1},...,s^{1}_{m}$ and be equal to $0$ at
all other edges of the graph $G_{3}$. The correlation function at the 
edges $s^{1}_{1},...,s^{1}_{m}$ of the lattice $G_{3}$ is the function
\begin{equation}
\label{2.12}
W_{G_{3}}(\chi^{1} ) = (Z_{G_{3}})^{- 1}
\sum_{A^{1} \in C^{1}(P(G_{3}),Z^{add}_{2})}
(- 1)^{\langle \chi^{1} ,A^{1}\rangle }
\exp \{ - \beta H(\partial^{\star} A^{1})\} .
\end{equation}
{\bf Proposition 2.1.} {\it The partition function of} ${\bf Z}_{2}$
{\it electrodynamics on the graph} $G_{3}$
\begin{equation}
\label{2.13}
Z_{G_{3}} = 2^{\# (EG_{3})}
\left( \prod_{s^{2}_{i} \in P(G_{3})} \cosh \beta E(s^{2}_{i})\right)
Z_{r,G_{3}}
\end{equation}
{\it where the reduced partition function of} ${\bf Z}_{2}$
{\it electrodynamics on the graph} $G_{3}$
\begin{equation}
\label{2.14}
Z_{r,G_{3}} = \sum_{\xi^{2} \in Z_{2}(P(G_{3}),Z^{add}_{2})}
\prod_{s^{2}_{i} \in P(G_{3})}
(\tanh \beta E(s^{2}_{i}))^{\frac{1}{2} (1 - (- 1)^{\xi^{2} (s^{2}_{i})})}
\end{equation}
{\it and} $\# (EG_{3})$ {\it is the total number of the non -- oriented
edges of the graph} $G_{3}$.

{\it The correlation function of} ${\bf Z}_{2}$ {\it electrodynamics
on the graph} $G_{3}$
\begin{equation}
\label{2.15}
W_{G_{3}}(\chi^{1} ) = (Z_{r,G_{3}})^{- 1}
\sum_{{\xi^{2} \in C^{2}(P(G_{3}),Z^{add}_{2}),} \atop 
{\partial \xi^{2} \, = \, \chi^{1}}}
\prod_{s^{2}_{i} \in P(G_{3})}
(\tanh \beta E(s^{2}_{i}))^{\frac{1}{2} (1 - (- 1)^{\xi^{2} (s^{2}_{i})})}.
\end{equation}
{\it Proof.} The definition (\ref{2.10}) implies
\begin{equation}
\label{2.16}
\exp \{ - \beta H(A^{2})\} = \prod_{s^{2}_{i} \in P(G_{3})}
\exp \{ \beta E(s^{2}_{i})(- 1)^{A^{2}(s^{2}_{i})}\}
\end{equation}
where $A^{2} \in C^{2}(P(G_{3}),{\bf Z}^{add}_{2})$. It is easy to
verify that for $\epsilon = 0,1$
\begin{equation}
\label{2.17}
\exp \{ \beta E(s^{2}_{i})(- 1)^{\epsilon }\} =
(\cosh \beta E(s^{2}_{i}))\sum_{\xi = 0,1}
(- 1)^{\xi \epsilon }
(\tanh \beta E(s^{2}_{i}))^{\frac{1}{2} (1 - (- 1)^{\xi })}.
\end{equation}
The relations (\ref{2.16}), (\ref{2.17}) imply
\begin{eqnarray}
\label{2.18}
\exp \{ - \beta H(A^{2})\} = \nonumber \\
\left( \prod_{s^{2}_{i} \in P(G_{3})} \cosh \beta E(s^{2}_{i})\right)
\sum_{\xi^{2} \in C^{2}(P(G_{3}),Z^{add}_{2})}
(- 1)^{\langle \xi^{2}, A^{2}\rangle }
\prod_{s^{2}_{i} \in P(G_{3})}
(\tanh \beta E(s^{2}_{i}))^{\frac{1}{2} (1 - (- 1)^{\xi^{2} (s^{2}_{i})})}.
\end{eqnarray}
The substitution of the equality (\ref{2.18}) into the definition 
(\ref{2.11}), the first relation (\ref{2.6}) and the relation
\begin{equation}
\label{2.19}
\sum_{\xi = 0,1} (- 1)^{\xi \epsilon} =
\left\{ {2, \hskip 1cm \epsilon = 0,} \atop
{0, \hskip 1cm \epsilon = 1,} \right.
\end{equation}
give the equalities (\ref{2.13}), (\ref{2.14}). The substitution of the
equality (\ref{2.18}) into the definition (\ref{2.12}), the first relation
(\ref{2.6}) and the relation (\ref{2.19}) give the equality (\ref{2.15}).
The proposition is proved.

Here we used the definitions and the methods of the paper \cite{7}.

The relation (2.15) implies that the correlation function is not zero only 
for the cochain $\chi^{1} \in B_{1}(P(G_{3}),{\bf Z}^{add}_{2})$. In other
words, the cochain $\chi^{1} $ is equal to $1$ on the closed contours
which are the boundaries of the two -- dimensional domains.

Let us consider a rectangular lattice formed by the points with the 
integral Cartesian coordinates $x = k_{1}$, $y = k_{2}$,
$M^{\prime}_{i} \leq k_{i} \leq M_{i}$, $i = 1,2$, and the corresponding
edges connecting neighbour vertices. We denote this graph by 
$G_{2} = G(M^{\prime}_{1},M^{\prime}_{2};M_{1},M_{2})$. The cell complex
$P(G_{2})$ is defined analogously to the cell complex $P(G_{3})$ for
the three -- dimensional graph $G_{3} =
G(M^{\prime}_{1},M^{\prime}_{2},M^{\prime}_{3};M_{1},M_{2},M_{3})$.
Let a cochain $\sigma^{0} \in C^{0}(P(G_{2}),{\bf Z}^{add}_{2})$.
The Ising model energy is expressed in the following form similar to
the form (\ref{2.10})
\begin{equation}
\label{2.20}
H(\partial^{\star} \sigma^{0} ) = - \sum_{s^{1}_{i} \in P(G_{2})}
E(s^{1}_{i})(- 1)^{\partial^{\star} \sigma^{0} (s^{1}_{i})}.
\end{equation}
The partition function and the correlation function of the 
two -- dimensional Ising model are defined by the relations similar to
the relations (\ref{2.11}), (\ref{2.12})
\begin{equation}
\label{2.21}
Z_{G_{2}} = \sum_{\sigma^{0} \in C^{0}(P(G_{2}),Z^{add}_{2})}
\exp \{ - \beta H(\partial^{\star} \sigma^{0} )\},
\end{equation}
\begin{equation}
\label{2.22}
W_{G_{2}}(\chi^{0} ) = (Z_{G_{2}})^{- 1}
\sum_{\sigma^{0} \in C^{0}(P(G_{2}),Z^{add}_{2})}
(- 1)^{\langle \chi^{0} ,\sigma^{0} \rangle }
\exp \{ - \beta H(\partial^{\star} \sigma^{0} )\} .
\end{equation}
where the cochain $\chi^{0} \in C^{0}(P(G_{2}),{\bf Z}^{add}_{2})$.

It is possible to show \cite{6} that similar to the relations
(\ref{2.13}) -- (\ref{2.15})
\begin{equation}
\label{2.23}
Z_{G_{2}} = 2^{\# (VG_{2})}
\left( \prod_{s^{1}_{i} \in P(G_{2})} \cosh \beta E(s^{1}_{i})\right)
Z_{r,G_{2}},
\end{equation}
\begin{equation}
\label{2.24}
Z_{r,G_{2}} = \sum_{\xi^{1} \in Z_{1}(P(G_{2}),Z^{add}_{2})}
\prod_{s^{1}_{i} \in P(G_{2})}
(\tanh \beta E(s^{1}_{i}))^{\frac{1}{2} (1 - (- 1)^{\xi^{1} (s^{1}_{i})})},
\end{equation}
\begin{equation}
\label{2.25}
W_{G_{2}}(\chi^{0} ) = (Z_{r,G_{2}})^{- 1}
\sum_{{\xi^{1} \in C^{1}(P(G_{2}),Z^{add}_{2}),} \atop 
{\partial \xi^{1} \, = \, \chi^{0}}}
\prod_{s^{1}_{i} \in P(G_{2})}
(\tanh \beta E(s^{1}_{i}))^{\frac{1}{2} (1 - (- 1)^{\xi^{1} (s^{1}_{i})})}
\end{equation}
where $\# (VG_{2})$ is the total number of the vertices of the graph
$G_{2}$.

The equalities (\ref{2.23}), (\ref{2.24}) were proved for the first time
in the paper \cite{8}.

Let us consider the relations (\ref{2.13}) -- (\ref{2.15}) for the
particular case of the interaction energy $E(s^{2}_{i}) = 0$ for any face
$s^{2}_{i}$ orthogonal to the coordinate axis $z$. For the graph $G_{3} =
G(M^{\prime}_{1},M^{\prime}_{2},M^{\prime}_{3};M_{1},M_{2},M_{3})$
the group $Z_{2}(P(G_{3}),{\bf Z}^{add}_{2})$ coincides with the group

\noindent
$B_{2}(P(G_{3}),{\bf Z}^{add}_{2})$. (The homology group is trivial).
A boundary of any set of the cubes contains a face $s^{2}_{i}$
orthogonal to the coordinate axis $z$. Hence the equality $E(s^{2}_{i}) = 0$
implies that the sum (\ref{2.14}) consists of the only term corresponding
to the cycle which is equal to zero on any face. Thus the relations
(\ref{2.13}), (\ref{2.14}) imply
\begin{equation}
\label{2.26}
Z_{G_{3}} = 2^{\# (EG_{3})}
\prod_{s^{2}_{i} \in P(G_{3})} \cosh \beta E(s^{2}_{i}).
\end{equation}
Let us consider the sum (\ref{2.15}). It is possible that the equation
$\partial \xi^{2} = \chi^{1} $ has no solution 
$\xi^{2} \in C^{2}(P(G_{3}),{\bf Z}^{add}_{2})$ such that the cochain
$\xi^{2} (s^{2}_{i}) = 0$ for any face $s^{2}_{i}$ orthogonal to the
coordinate axis $z$. In this case the relation (\ref{2.15}) implies
\begin{equation}
\label{2.27}
W_{G_{3}} (\chi^{1} ) = 0.
\end{equation}
If the equation $\partial \xi^{2} = \chi^{1} $ has a solution
$\xi^{2} \in C^{2}(P(G_{3}),{\bf Z}^{add}_{2})$ such that the cochain
$\xi^{2} (s^{2}_{i}) = 0$ for any face $s^{2}_{i}$ orthogonal to the
coordinate axis $z$, then this solution is unique since the equality
$\xi^{2} (s^{2}_{i}) = 0$ for any face $s^{2}_{i}$ orthogonal to the
coordinate axis $z$ implies that the cycle
$\xi^{2} \in Z_{2}(P(G_{3}),{\bf Z}^{add}_{2})$ is equal to zero on any
face. It follows from the relations (\ref{2.13}) -- (\ref{2.15}), 
(\ref{2.26}) that in this case
\begin{equation}
\label{2.28}
W_{G_{3}} (\chi^{1} ) = 
\prod_{{s^{2}_{i} \in P(G_{3}),} \atop {\partial \xi^{2} = \chi^{1}}}
(\tanh \beta E(s^{2}_{i}))^{\frac{1}{2} (1 - (- 1)^{\xi^{2} (s^{2}_{i})})}.
\end{equation}

It is possible to set the infinite interaction energies 
$E(s^{2}_{i}) = \pm \infty $ into the reduced partition function 
(\ref{2.14}) and the correlation functions (\ref{2.15}).

\section{Partition Function}
\setcounter{equation}{0}

\noindent In this section we study the connection between the reduced
partition function (\ref{2.14}) of the three -- dimensional ${\bf Z}_{2}$
electrodynamics and the reduced partition function (\ref{2.24}) of the
two -- dimensional Ising model.

We consider the finite three -- dimensional lattice
$G(M^{\prime}_{1},M^{\prime}_{2},M^{\prime}_{3};M_{1},M_{2},M_{3})$.
The non -- oriented edge of this lattice is given by the pair 
$\{ {\bf p},{\bf e}\} $ where ${\bf p}$ is the edge initial vertex and 
the unit vector ${\bf e}$ is the edge direction. The unit vector 
${\bf e}$ is one of six vectors: $(\pm 1,0,0)$, $(0,\pm 1,0)$, $(0,0,\pm 1)$. 
Since the edge is non -- oriented, then 
$\{ {\bf p} + {\bf e}, - {\bf e}\} = \{ {\bf p},{\bf e}\} $. The intitial
vertex ${\bf p}$ and the final vertex ${\bf p} + {\bf e}$ of the edge
$\{ {\bf p},{\bf e}\} $ are the vertices of the graph
$G(M^{\prime}_{1},M^{\prime}_{2},M^{\prime}_{3};M_{1},M_{2},M_{3})$. 
Hence the components $p_{i}, i = 1,2,3$, are the natural numbers and
$M^{\prime}_{i} \leq p_{i} \leq M_{i}$, 
$M^{\prime}_{i} \leq p_{i} + e_{i} \leq M_{i}$, $i = 1,2,3$. For the edge
$\{ {\bf p},{\bf e}\} $ the incidence numbers $(\{ {\bf p},{\bf e}\} :{\bf p}) 
= (\{ {\bf p},{\bf e}\} :{\bf p} + {\bf e}) = 1$. All other incidence
numbers are equal to zero.

The non -- oriented face of the lattice 
$G(M^{\prime}_{1},M^{\prime}_{2},M^{\prime}_{3};M_{1},M_{2},M_{3})$
is given by the triplet $\{ {\bf p},{\bf e}_{1},{\bf e}_{2}\} $ 
where ${\bf p}$ is the face initial vertex and the unit vectors 
${\bf e}_{1},{\bf e}_{2}$ are two orthogonal to each other vectors from 
six unit vectors $(\pm 1,0,0)$, $(0,\pm 1,0)$, $(0,0,\pm 1)$. Since the 
face is non -- oriented, then 
$\{ {\bf p},{\bf e}_{1},{\bf e}_{2}\} = \{ {\bf p},{\bf e}_{2},{\bf e}_{1}\} =
\{ {\bf p} + {\bf e}_{1}, - {\bf e}_{1},{\bf e}_{2}\} =
\{ {\bf p} + {\bf e}_{2},{\bf e}_{1}, - {\bf e}_{2}\} =
\{ {\bf p} + {\bf e}_{1} + {\bf e}_{2}, - {\bf e}_{1}, - {\bf e}_{2}\}  $.
Any of four vertices ${\bf p}, {\bf p} + {\bf e}_{1}, {\bf p} + {\bf e}_{2},
{\bf p} + {\bf e}_{1} + {\bf e}_{2}$ of the face
$\{ {\bf p},{\bf e}_{1},{\bf e}_{2}\} $ is the vertex of the graph
$G(M^{\prime}_{1},M^{\prime}_{2},M^{\prime}_{3};M_{1},M_{2},M_{3})$.
Hence the components $p_{i}, i = 1,2,3$, are the natural numbers and
$M^{\prime}_{i} \leq p_{i} \leq M_{i}, 
M^{\prime}_{i} \leq p_{i} + ({\bf e}_{1})_{i} \leq M_{i},
M^{\prime}_{i} \leq p_{i} + ({\bf e}_{2})_{i} \leq M_{i},
M^{\prime}_{i} \leq p_{i} + ({\bf e}_{1})_{i} + ({\bf e}_{2})_{i} \leq M_{i}, 
i = 1,2,3$. For the non -- oriented face 
$\{ {\bf p},{\bf e}_{1},{\bf e}_{2}\} $ the incidence numbers
$(\{ {\bf p},{\bf e}_{1},{\bf e}_{2}\} :\{ {\bf p},{\bf e}_{1}\} ) =
(\{ {\bf p},{\bf e}_{1},{\bf e}_{2}\} :\{ {\bf p},{\bf e}_{2}\}) =
(\{ {\bf p},{\bf e}_{1},{\bf e}_{2}\} :\{ {\bf p} + {\bf e}_{2},{\bf e}_{1}\})
= 
(\{ {\bf p},{\bf e}_{1},{\bf e}_{2}\} :\{ {\bf p} + {\bf e}_{1},{\bf e}_{2}\})
= 1$. All other incidence numbers are equal to zero.

\noindent {\bf Theorem 3.1.} {\it Let the reduced partition function}
$Z_{r,G_{3}}$ {\it of the three -- dimensional} ${\bf Z}_{2}$
{\it electrodynamics be given by the relation} (\ref{2.14}). {\it Let the
reduced partition function} $Z_{r,G_{2}}$ {\it of the two -- dimensional
Ising model be given by the relation} (\ref{2.24}). {\it Then}
\begin{eqnarray}
\label{3.1}
Z_{r,G(M^{\prime}_{1},M^{\prime}_{2},M^{\prime}_{3};M_{1},M_{2},M_{3})}
|_{E(\{ p,(1,0,0),(0,1,0)\} ) \, = \, \infty }  = \nonumber \\
Z_{r,G(M^{\prime}_{1},M^{\prime}_{2},M^{\prime}_{3};M_{1},M_{2},M_{3})}
|_{E(\{ p,(1,0,0),(0,1,0)\} ) \, = \, - \, \infty } = \nonumber \\
\prod^{M_{3} - 1}_{i \, = \, M^{\prime}_{3}}
Z_{r,G(M^{\prime}_{1},M^{\prime}_{2};M_{1},M_{2}),i}
\end{eqnarray}
{\it where} $\{ {\bf p},(1,0,0),(0,1,0)\} $ {\it is any face orthogonal 
to the coordinate axis} $z$. {\it The reduced partition function}
$Z_{r,G(M^{\prime}_{1},M^{\prime}_{2};M_{1},M_{2}),i}$ {\it is defined
by the relation} (\ref{2.24}) {\it with the interaction energies}
$E(\{ (p_{1},p_{2}),{\bf e}\} ) = E(\{ (p_{1},p_{2},i),{\bf e}, (0,0,1)\} )$.

\noindent {\it Proof.} Let us consider a cycle
$\xi^{2} \in Z_{2}(P(G_{3}),{\bf Z}^{add}_{2})$ for the graph
$G_{3} = $

\noindent
$G(M^{\prime}_{1},M^{\prime}_{2},M^{\prime}_{3};M_{1},M_{2},M_{3})$.
By using this cycle we define $M_{3} - M^{\prime}_{3}$ cochains from the
group $C^{1}(P(G_{2}),{\bf Z}^{add}_{2})$ for the graph 
$G_{2} = G(M^{\prime}_{1},M^{\prime}_{2};M_{1},M_{2})$
\begin{equation}
\label{3.2}
\xi^{1}_{i} (\{ (p_{1},p_{2}),{\bf e}\} ) =
\xi^{2} (\{ (p_{1},p_{2},i),{\bf e},(0,0,1)\} )
\end{equation}
where $i = M^{\prime}_{3},...,M_{3} - 1$ and the unit vector ${\bf e}$
is orthogonal to the unit vector $(0,0,1)$. The incidence numbers
satisfy the following relations
\begin{equation}
\label{3.3}
(\{ (p_{1},p_{2}),{\bf e}\} :(q_{1},q_{2})) =
(\{ (p_{1},p_{2},i),{\bf e},(0,0,1)\} :\{ (q_{1},q_{2},i),(0,0,1)\} )
\end{equation}
where $i = M^{\prime}_{3},...,M_{3} - 1$. Hence the definition (\ref{2.3})
implies
\begin{equation}
\label{3.4}
\partial \xi^{1}_{i} ((p_{1},p_{2})) =
\partial \xi^{2} (\{ (p_{1},p_{2},i),(0,0,1)\} ).
\end{equation}
Since $\xi^{2} \in Z_{2}(P(G_{3}),{\bf Z}^{add}_{2})$ the equality 
(\ref{3.4}) implies $\xi^{1}_{i} \in Z_{1}(P(G_{2}),{\bf Z}^{add}_{2})$.

By using a cycle $\xi^{2} \in Z_{2}(P(G_{3}),{\bf Z}^{add}_{2})$ we
define $M_{3} - M^{\prime}_{3} + 1$ cochains from the group
$C^{2}(P(G_{2}),{\bf Z}^{add}_{2})$ for the graph 
$G_{2} = G(M^{\prime}_{1},M^{\prime}_{2};M_{1},M_{2})$
\begin{equation}
\label{3.5}
\xi^{2}_{i} (\{ (p_{1},p_{2}),{\bf e}_{1},{\bf e}_{2}\} ) =
\xi^{2} (\{ (p_{1},p_{2},i),{\bf e}_{1},{\bf e}_{2}\} )
\end{equation}
where $i = M^{\prime}_{3},...,M_{3}$ and the unit vectors 
${\bf e}_{1},{\bf e}_{2}$ are orthogonal to the unit vector $(0,0,1)$.
Since $\xi^{2} \in Z_{2}(P(G_{3}),{\bf Z}^{add}_{2})$ the definition
(\ref{2.3}) and the relations (\ref{3.2}), (\ref{3.5}) imply
\begin{equation}
\label{3.6}
\partial \xi^{2}_{i} (\{ (p_{1},p_{2}),{\bf e}\} ) =
\xi^{1}_{i - 1} (\{ (p_{1},p_{2}),{\bf e}\} ) +
\xi^{1}_{i} (\{ (p_{1},p_{2}),{\bf e}\} ),
\end{equation}
$i = M^{\prime}_{3} + 1,...,M_{3} - 1$,
\begin{equation}
\label{3.7}
\partial \xi^{2}_{M^{\prime}_{3}} (\{ (p_{1},p_{2}),{\bf e}\} ) =
\xi^{1}_{M^{\prime}_{3}} (\{ (p_{1},p_{2}),{\bf e}\} ),
\end{equation}
\begin{equation}
\label{3.8}
\partial \xi^{2}_{M_{3}} (\{ (p_{1},p_{2}),{\bf e}\} ) =
\xi^{1}_{M_{3} - 1} (\{ (p_{1},p_{2}),{\bf e}\} )
\end{equation}
where the unit vector ${\bf e}$ is orthogonal to the unit vector $(0,0,1)$.

For the graph $G_{2} = G(M^{\prime}_{1},M^{\prime}_{2};M_{1},M_{2})$
the group $Z_{1}(P(G_{2}),{\bf Z}^{add}_{2})$ coincides with the group
$B_{1}(P(G_{2}),{\bf Z}^{add}_{2})$ and the group
$Z_{2}(P(G_{2}),{\bf Z}^{add}_{2})$ consists of the only cochain which
is equal to zero on any face. (The homology groups are trivial). Let the
arbitrary cycles $\xi^{1}_{i} \in Z_{1}(P(G_{2}),{\bf Z}^{add}_{2})$,
$i = M^{\prime}_{3},...,M_{3} - 1$, be given. Then the equations
(\ref{3.6}) -- (\ref{3.8}) define the cochains $\xi^{2}_{i}$,
$i = M^{\prime}_{3},...,M_{3}$, uniquely. Now the relations (\ref{3.2}),
(\ref{3.5}) define uniquely the cycle
$\xi^{2} \in Z_{2}(P(G_{3}),{\bf Z}^{add}_{2})$. Hence the cycle
$\xi^{2} \in Z_{2}(P(G_{3}),{\bf Z}^{add}_{2})$ is given uniquely by the
cycles $\xi^{1}_{i} \in Z_{1}(P(G_{2}),{\bf Z}^{add}_{2})$,
$i = M^{\prime}_{3},...,M_{3} - 1$. The relations (\ref{3.2}), (\ref{3.5})
imply
\begin{eqnarray}
\label{3.9}
\prod_{\{ p,e_{1},e_{2}\} \in P(G_{3})}
(\tanh \beta E(\{ {\bf p},{\bf e}_{1},{\bf e}_{2}\} ))^{\frac{1}{2} 
(1 - (- 1)^{\xi^{2} (\{ p,e_{1},e_{2}\} )})} = \nonumber \\
\left( \prod^{M_{3}}_{i \, = \, M^{\prime}_{3}}
\prod_{\{ (p_{1},p_{2}),e_{1},e_{2}\} \in P(G_{2})}
(\tanh \beta E(\{ (p_{1},p_{2},i),{\bf e}_{1},{\bf e}_{2}\} ))^{\frac{1}{2} 
(1 - (- 1)^{\xi^{2}_{i} (\{ (p_{1},p_{2}),e_{1},e_{2}\} )})} \right)
\times \nonumber \\
\left( \prod^{M_{3} - 1}_{i \, = \, M^{\prime}_{3}}
\prod_{\{ (p_{1},p_{2}),e\} \in P(G_{2})}
(\tanh \beta E(\{ (p_{1},p_{2},i),{\bf e},(0,0,1)\} ))^{\frac{1}{2} 
(1 - (- 1)^{\xi^{1}_{i} (\{ (p_{1},p_{2}),e\} )})} \right)
\end{eqnarray} 
Since $\beta > 0$, then
\begin{equation}
\label{3.10}
(\tanh \beta E(\{ {\bf p},{\bf e}_{1},{\bf e}_{2}\} ))
|_{E(\{ p,e_{1},e_{2}\} ) \, = \, \pm \, \infty } = \pm 1.
\end{equation}
If $\epsilon = \pm 1$, then for any natural number $m$ we have
\begin{equation}
\label{3.11}
\epsilon^{\frac{1}{2} (1 - (- 1)^{m})} = \epsilon^{m}.
\end{equation}
It follows from the relations (\ref{3.9}) -- (\ref{3.11})
\begin{eqnarray}
\label{3.12}
\left( \prod_{\{ p,e_{1},e_{2}\} \in P(G_{3})}
(\tanh \beta E(\{ {\bf p},{\bf e}_{1},{\bf e}_{2}\} ))^{\frac{1}{2} 
(1 - (- 1)^{\xi^{2} (\{ p,e_{1},e_{2}\} )})}
\right)|_{E(\{ p,(1,0,0),(0,1,0)\} ) \, = \, \pm \, \infty }
= \nonumber \\
\left( \prod_{\{ (p_{1},p_{2}),e_{1},e_{2}\} \in P(G_{2})}
\epsilon^{ \sum^{M_{3}}_{i \, = \, M^{\prime}_{3}} \,
\xi^{2}_{i} (\{ (p_{1},p_{2}),e_{1},e_{2}\} )} \right)
\times \nonumber \\
\left( \prod^{M_{3} - 1}_{i \, = \, M^{\prime}_{3}}
\prod_{\{ (p_{1},p_{2}),e\} \in P(G_{2})}
(\tanh \beta E(\{ (p_{1},p_{2},i),{\bf e},(0,0,1)\} ))^{\frac{1}{2} 
(1 - (- 1)^{\xi^{1}_{i} (\{ (p_{1},p_{2}),e\} )})} \right)
\end{eqnarray}
where $E(\{ {\bf p},(1,0,0),(0,1,0)\} )$ is an interaction energy
attached to a face 

\noindent $\{ {\bf p},(1,0,0),(0,1,0)\} $ orthogonal 
to the coordinate axis $z$ and $\epsilon = \pm 1$ in correspondence
with the relation (\ref{3.10}).

By summing up the relations (\ref{3.6}) -- (\ref{3.8}) we have
\begin{equation}
\label{3.13}
\partial \left( \sum^{M_{3}}_{i \, = \, M^{\prime}_{3}} \xi^{2}_{i} \right)
= 0.
\end{equation}
The group $Z_{2}(P(G_{2}),{\bf Z}^{add}_{2})$ consists of the only 
cochain which is equal to zero on any face. (The homology group is trivial).
Hence the relations (\ref{2.14}), (\ref{2.24}), (\ref{3.12}) and
(\ref{3.13}) imply the equality (\ref{3.1}). The theorem is proved.

Let us consider the particular case when the interaction energy depends
only on the face orientation: $E(\{ {\bf p},(1,0,0),(0,0,1)\} ) = E_{1}$,
$E(\{ {\bf p},(0,1,0),(0,0,1)\} ) = E_{2}$. Then the relation (\ref{3.1})
implies
\begin{eqnarray}
\label{3.14}
Z_{r,G(M^{\prime}_{1},M^{\prime}_{2},M^{\prime}_{3};M_{1},M_{2},M_{3})}
|_{E(\{ p,(1,0,0),(0,1,0)\} ) \, = \, \infty }  = \nonumber \\
Z_{r,G(M^{\prime}_{1},M^{\prime}_{2},M^{\prime}_{3};M_{1},M_{2},M_{3})}
|_{E(\{ p,(1,0,0),(0,1,0)\} ) \, = \, - \, \infty } = \nonumber \\
(Z_{r,G(M^{\prime}_{1},M^{\prime}_{2};M_{1},M_{2})})^{M_{3} - M^{\prime}_{3}}
\end{eqnarray}
where the reduced partition function 
$Z_{r,G(M^{\prime}_{1},M^{\prime}_{2};M_{1},M_{2})}$ is defined by the
relation (\ref{2.24}) for the interaction energies 
$E((p_{1},p_{2}),(1,0)) = E_{1}$, $E((p_{1},p_{2}),(0,1)) = E_{2}$.

The total number $\# (EG_{3})$ of the non -- oriented edges of the graph
$G_{3} = $

\noindent
$G(M^{\prime}_{1},M^{\prime}_{2},M^{\prime}_{3};M_{1},M_{2},M_{3})$
is equal to
$$
(M_{1} - M^{\prime}_{1} + 1)(M_{2} - M^{\prime}_{2} + 1)
(M_{3} - M^{\prime}_{3}) +
(M_{1} - M^{\prime}_{1} + 1)(M_{2} - M^{\prime}_{2})
(M_{3} - M^{\prime}_{3} + 1) +
$$
$$
(M_{1} - M^{\prime}_{1})(M_{2} - M^{\prime}_{2} + 1)
(M_{3} - M^{\prime}_{3} + 1).
$$
Hence the relation (\ref{3.14}) and Theorem 4.2 from the paper \cite{6}
imply
\begin{eqnarray}
\label{3.15}
\lim_{G_{3} \rightarrow Z^{\times 3}} (\# (EG_{3}))^{- 1}
(\ln Z_{r,G_{3}})|_{E(\{ p,(1,0,0),(0,1,0)\} ) \, = \, \infty }  = 
\nonumber \\
\lim_{G_{3} \rightarrow Z^{\times 3}} (\# (EG_{3}))^{- 1}
(\ln Z_{r,G_{3}})|_{E(\{ p,(1,0,0),(0,1,0)\} ) \, = \, - \, \infty }  = 
\nonumber \\
\frac{1}{3} \lim_{{M_{i} \, \rightarrow \, \infty, \, 
M^{\prime}_{i} \, \rightarrow \, - \, \infty,} \atop {i \, = \, 1,2}}
(M_{1} - M^{\prime}_{1} + 1)^{- 1}(M_{2} - M^{\prime}_{2} + 1)^{- 1}
\ln Z_{r,G(M^{\prime}_{1},M^{\prime}_{2};M_{1},M_{2})} = \nonumber \\
\frac{1}{6} (2\pi )^{- 2} \int^{2\pi }_{0} d\theta_{1} 
\int^{2\pi }_{0} d\theta_{2} \nonumber \\
\ln [(1 + z^{2}_{1})(1 + z^{2}_{2}) - 2z_{1}(1 - z^{2}_{2})\cos \theta_{1}
- 2z_{2}(1 - z^{2}_{1})\cos \theta_{2} ]
\end{eqnarray} 
where the variables $z_{i} = \tanh \beta E_{i}$, $i = 1,2$, satisfy the
estimate $|z_{i}| < 1/3$.

It is easy to show that the integral (\ref{3.15}) has the singularity
when
\begin{equation}
\label{3.16}
|z_{1}||z_{2}| + |z_{1}| + |z_{2}| = 1.
\end{equation}

\section{Correlation Functions}
\setcounter{equation}{0}

\noindent In this section we study the connection between the correlation
functions of the three -- dimensional ${\bf Z}_{2}$ electrodynamics
and the correlation functions of the two -- dimensional Ising model.

For the fixed cochain 
$\chi^{0} \in C^{0}(P(G_{2}(\chi^{0} )),{\bf Z}^{add}_{2})$ where the
graph $G_{2}(\chi^{0} ) = $

\noindent $G(M^{\prime}_{1}(\chi^{0} ),
M^{\prime}_{2}(\chi^{0} );M_{1}(\chi^{0} ),M_{2}(\chi^{0} ))$ we define
\begin{equation}
\label{4.1}
W_{Z^{\times 2}}(\chi^{0} ) =
\lim_{{G_{2} \, \rightarrow \, Z^{\times 2},} \atop 
{G_{2}(\chi^{0} ) \, \subset \, G_{2}}}
W_{G_{2}}(\chi^{0} ).
\end{equation}
Due to \cite{6} we describe the correlation function (\ref{4.1}). The
oriented edge of the graph ${\bf Z}^{\times 2}$ is giveh by the pair
$({\bf p},{\bf e})$ where ${\bf p}$ is the edge initial vertex and the
unit vector ${\bf e}$ is the edge direction. The unit vector ${\bf e}$
is one of four vectors: $(\pm 1,0)$, $(0,\pm 1)$. A closed path is a
sequence of the oriented edges 
$C = (({\bf p}_{1},{\bf e}_{1}),...,({\bf p}_{k},{\bf e}_{k}))$ such that
\begin{equation}
\label{4.2}
{\bf p}_{i + 1} = {\bf p}_{i} + {\bf e}_{i}, \, i = 1,...,k - 1, \,
{\bf p}_{1} = {\bf p}_{k} + {\bf e}_{k}.
\end{equation}
The number $|C| = k$ is the length of the closed path
$C = (({\bf p}_{1},{\bf e}_{1}),...,({\bf p}_{k},{\bf e}_{k}))$.
If $C$ is a closed path, the support $||C||$ is the set of all 
non -- oriented edges $\{ {\bf p},{\bf e}\} $ such that the oriented
edge $({\bf p},{\bf e})$ is included into the path $C$ or the oriented
edge $({\bf p} + {\bf e}, - {\bf e})$ is included into the path $C$.
The closed path $(({\bf p}_{1},{\bf e}_{1}),...,({\bf p}_{k},{\bf e}_{k}))$
is called reduced if it satisfies the following condition
\begin{equation}
\label{4.3}
{\bf p}_{i + 1} = {\bf p}_{i} + {\bf e}_{i}, \, 
{\bf p}_{i + 1} + {\bf e}_{i + 1} \neq {\bf p}_{i}, \, i = 1,...,k - 1, \,
{\bf p}_{1} = {\bf p}_{k} + {\bf e}_{k}, \, 
{\bf p}_{1} + {\bf e}_{1} \neq {\bf p}_{k}.
\end{equation}
The set of all reduced closed paths on the graph ${\bf Z}^{\times 2}$
is denoted by $RC({\bf Z}^{\times 2})$.

Let with any pair $({\bf p}_{1},{\bf e}_{1})$, $({\bf p}_{2},{\bf e}_{2})$
of the oriented edges of the graph ${\bf Z}^{\times 2}$ such that
${\bf p}_{2} = {\bf p}_{1} + {\bf e}_{1}$, 
${\bf p}_{2} + {\bf e}_{2} \neq {\bf p}_{1}$ there correspond the number
\begin{equation}
\label{4.4}
\rho (({\bf p}_{1},{\bf e}_{1});({\bf p}_{2},{\bf e}_{2})) =
\exp \{ \frac{i}{2} \hat{({\bf e}_{1},{\bf e}_{2})}\}
\end{equation}
where $\hat{({\bf e}_{1},{\bf e}_{2})}$ is the radian measure of the
angle between the direction of the vector ${\bf e}_{1}$ and the direction
of the vector ${\bf e}_{2}$. With any reduced closed path
$C = (({\bf p}_{1},{\bf e}_{1}),...,({\bf p}_{k},{\bf e}_{k}))$ on the
graph ${\bf Z}^{\times 2}$ there corresponds the number
\begin{eqnarray}
\label{4.5}
\rho (C) = \rho (({\bf p}_{1},{\bf e}_{1});({\bf p}_{2},{\bf e}_{2}))
\rho (({\bf p}_{2},{\bf e}_{2});({\bf p}_{3},{\bf e}_{3})) \cdots
\rho (({\bf p}_{k - 1},{\bf e}_{k - 1});({\bf p}_{k},{\bf e}_{k})) \times
\nonumber \\
\rho (({\bf p}_{k},{\bf e}_{k});({\bf p}_{1},{\bf e}_{1})).
\end{eqnarray}
The number $\rho (C) = \exp \{ \frac{i}{2} \phi (C)\} $ where $\phi (C)$
is the total angle through which the tangent vector of the path $C$
turns along the path $C$. 

Let a cochain $\xi^{1} \in C^{1}(P({\bf Z}^{\times 2}),{\bf Z}^{add}_{2})$.
The support $||\xi^{1} ||$ is the set of all non -- oriented edges of the
graph ${\bf Z}^{\times 2}$ on which a cochain $\xi^{1} $ takes the value
$1$. Let a cochain 
$\chi^{0} \in C^{0}(P({\bf Z}^{\times 2}),{\bf Z}^{add}_{2})$. The
support $||\xi^{1} ||$ is called $\chi^{0} $ -- connected if any 
connected component of the support $||\xi^{1} ||$ contains the 
non -- oriented edges incident to the vertices on which a cochain
$\chi^{0} $ equals $1$. Let $i(||\xi^{1} ||)$ be the set of all 
non -- oriented edges incident to the vertices incident to the edges
of the support $||\xi^{1} ||$.

Due to \cite{6} the correlation function (\ref{4.1}) has the following
form
\begin{eqnarray}
\label{4.6}
W_{Z^{\times 2}}(\chi^{0} ) =
\sum_{{\xi^{1} \, \in \, C^{1}(P(Z^{\times 2}),Z^{add}_{2}), 
\, \partial \xi^{1} \, = \, \chi^{0}, } \atop
{\chi^{0} \, - \, connected \, ||\xi^{1} ||}} \nonumber \\
\exp \{ \frac{1}{2} 
\sum_{{C \, \in \, RC(Z^{\times 2}),} \atop 
{||C|| \, \cap \, i(||\xi^{1} ||) \, \neq \, \emptyset }}
|C|^{- 1}\rho (C)\prod_{(p,e) \, \in \, C}
\tanh \beta E(\{ {\bf p},{\bf e}\})\} \times \nonumber \\
\prod_{\{ p,e\} \, \in \, P(Z^{\times 2})}
(\tanh \beta E(\{ {\bf p},{\bf e}\} ))^{\frac{1}{2} 
(1 - (- 1)^{\xi^{1} (\{ p,e\} )})}
\end{eqnarray}
The series (\ref{4.6}) is absolutely convergent when the following
estimate is valid
\begin{equation}
\label{4.7}
|\tanh \beta E(\{ {\bf p},{\bf e}\} )| < 1/3
\end{equation}
and the values of the interaction energies $E(\{ {\bf p},{\bf e}\} )$
have the same sign for all collinear vectors ${\bf e}$.

\noindent {\bf Theorem 4.1.} {\it Let for the graph} $G_{3}(\chi^{1} ) = $

\noindent $G(M^{\prime}_{1}(\chi^{1} ),M^{\prime}_{2}(\chi^{1} ),
M^{\prime}_{3}(\chi^{1} );M_{1}(\chi^{1} ),M_{2}(\chi^{1} ),M_{3}(\chi^{1}))$
{\it the cochain} $\chi^{1} \in B_{1}(P(G_{3}(\chi^{1})),{\bf Z}^{add}_{2})$
{\it be given. For the graph} $G_{2}(\chi^{1} ) = G(M^{\prime}_{1}(\chi^{1} ),
M^{\prime}_{2}(\chi^{1} );M_{1}(\chi^{1} ),M_{2}(\chi^{1} ))$ {\it we
define} $M_{3}(\chi^{1} ) - M^{\prime}_{3}(\chi^{1} )$ {\it cochains
from the group} $C^{0}(P(G_{2}(\chi^{1})),{\bf Z}^{add}_{2})$
\begin{equation}
\label{4.8}
\chi^{0}_{i} ((p_{1},p_{2})) = \chi^{1} (\{ (p_{1},p_{2},i),(0,0,1)\} ),
\end{equation}
$i = M^{\prime}_{3}(\chi^{1} ),...,M_{3}(\chi^{1} ) - 1$,
{\it and} $M_{3}(\chi^{1} ) - M^{\prime}_{3}(\chi^{1} ) + 1$ 
{\it cochains from the group} 

\noindent $C^{1}(P(G_{2}(\chi^{1})),{\bf Z}^{add}_{2})$
\begin{equation}
\label{4.9}
\chi^{1}_{i} (\{ (p_{1},p_{2}),{\bf e}\} ) = 
\chi^{1} (\{ (p_{1},p_{2}, i),{\bf e}\} )
\end{equation}
{\it where} $i = M^{\prime}_{3}(\chi^{1} ),...,M_{3}(\chi^{1} )$
{\it and the unit vector} ${\bf e}$ {\it is orthogonal to the unit
vector} $(0,0,1)$.

{\it If the interaction energies of the three -- dimensional} ${\bf Z}_{2}$
{\it electrodynamics satisfy the estimate}
\begin{equation}
\label{4.10}
|\tanh \beta E(\{ {\bf p},{\bf e},(0,0,1)\} )| < 1/3
\end{equation}
{\it and the values of the interaction energies}
$E(\{ {\bf p},{\bf e},(0,0,1)\} )$ {\it have the same sign for all 
collinear vectors} ${\bf e}$, {\it then the correlation functions of
the three -- dimensional} ${\bf Z}_{2}$ {\it electrodynamics}
\begin{eqnarray}
\label{4.11}
\lim_{{G_{3} \, \rightarrow \, Z^{\times 3},} \atop 
{G_{3}(\chi^{1} ) \, \subset \, G_{3}}}
W_{G_{3}}(\chi^{1} )|_{E(\{ p,(1,0,0),(0,1,0)\} ) \, = \, \infty }  
= \nonumber \\
\prod^{M_{3}(\chi^{1} ) \, - \, 1}_{i \, = \, M^{\prime}_{3}(\chi^{1} )}
W_{Z^{\times 2},i}(\chi^{0}_{i} ),
\end{eqnarray}
\begin{eqnarray}
\label{4.12}
\lim_{{G_{3} \, \rightarrow \, Z^{\times 3},} \atop 
{G_{3}(\chi^{1} ) \, \subset \, G_{3}}}
W_{G_{3}}(\chi^{1} )|_{E(\{ p,(1,0,0),(0,1,0)\} ) \, = \, - \, \infty }  
= \nonumber \\
(- 1)^{S(\sum^{M_{3}(\chi^{1} )}_{i \, = \, M^{\prime}_{3}(\chi^{1} )}
\chi^{1}_{i} )}
\prod^{M_{3}(\chi^{1} ) \, - \, 1}_{i \, = \, M^{\prime}_{3}(\chi^{1} )}
W_{Z^{\times 2},i}(\chi^{0}_{i} )
\end{eqnarray}
{\it where} $\{ {\bf p},(1,0,0),(0,1,0)\} $ {\it is a face orthogonal to
the coordinate axis} $z$. {\it The correlation function} 
$W_{Z^{\times 2},i}(\chi^{0} )$ {\it is given by the relation} (\ref{4.6}) 
{\it for the interaction energies}
$E(\{ (p_{1},p_{2}),{\bf e}\} ) = E(\{ (p_{1},p_{2},i),{\bf e},(0,0,1)\} )$.
{\it For any cochain} 
$\chi^{1} \in B_{1}(P(G_{3}(\chi^{1})),{\bf Z}^{add}_{2})$ {\it the
cochain}
$$
\sum^{M_{3}(\chi^{1} )}_{i \, = \, M^{\prime}_{3}(\chi^{1} )}
\chi^{1}_{i} = \partial \zeta^{2}
$$
{\it where the cochain} 
$\zeta^{2} \in C^{2}(P(G_{2}(\chi^{1})),{\bf Z}^{add}_{2})$
{\it is defined uniquely. Then}
\begin{equation}
\label{4.13}
(- 1)^{S(\sum^{M_{3}(\chi^{1} )}_{i \, = \, M^{\prime}_{3}(\chi^{1} )}
\chi^{1}_{i} )} =
(- 1)^{\sum_{\{ (p_{1},p_{2}),e_{1},e_{2}\} \, \in \, P(G_{2}(\chi^{1} ))}
\zeta^{2} (\{ (p_{1},p_{2}),e_{1},e_{2}\} )}.
\end{equation}
{\it Proof.} For the graph
$G_{3} = G(M^{\prime}_{1},M^{\prime}_{2},M^{\prime}_{3};M_{1},M_{2},M_{3}),
G_{3}(\chi^{1} ) \subset G_{3}$, we consider a cochain
$\xi^{2} \in C^{2}(P(G_{3}),{\bf Z}^{add}_{2})$ satisfying the equation
$\partial \xi^{2} = \chi^{1} $. The relations (\ref{3.2}) define the
cochains $\xi^{1}_{i} \in C^{1}(P(G_{2}),{\bf Z}^{add}_{2})$,
$i = M^{\prime}_{3},...,M_{3} - 1$, for the graph
$G_{2} = G(M^{\prime}_{1},M^{\prime}_{2};M_{1},M_{2})$. The relations 
(\ref{3.4}), (\ref{4.8}) imply
\begin{equation}
\label{4.14}
\partial \xi^{1}_{i} = \chi^{0}_{i}
\end{equation}
where $\chi^{0}_{i} = 0$ for $i < M^{\prime}_{3}(\chi^{1} )$ and
$i \geq M_{3}(\chi^{1} )$.

For the same cochain $\xi^{2} \in C^{2}(P(G_{3}),{\bf Z}^{add}_{2})$
the relations (\ref{3.5}) define the cochains
$\xi^{2}_{i} \in C^{2}(P(G_{2}),{\bf Z}^{add}_{2})$, 
$i = M^{\prime}_{3},...,M_{3}$. Since $\partial \xi^{2} = \chi^{1} $,
the definition (\ref{2.3}) and the relations (\ref{3.2}), (\ref{3.5}),
(\ref{4.9}) imply
\begin{equation}
\label{4.15}
\partial \xi^{2}_{i} (\{ (p_{1},p_{2}),{\bf e}\} ) = 
\chi^{1}_{i} (\{ (p_{1},p_{2}),{\bf e}\} ) +
\xi^{1}_{i - 1} (\{ (p_{1},p_{2}),{\bf e}\} ) +
\xi^{1}_{i} (\{ (p_{1},p_{2}),{\bf e}\} ),
\end{equation}
$i = M^{\prime}_{3} + 1,...,M_{3} - 1$,
\begin{equation}
\label{4.16}
\partial \xi^{2}_{M^{\prime}_{3}} (\{ (p_{1},p_{2}),{\bf e}\} ) = 
\chi^{1}_{M^{\prime}_{3}} (\{ (p_{1},p_{2}),{\bf e}\} ) +
\xi^{1}_{M^{\prime}_{3}} (\{ (p_{1},p_{2}),{\bf e}\} ),
\end{equation}
\begin{equation}
\label{4.17}
\partial \xi^{2}_{M_{3}} (\{ (p_{1},p_{2}),{\bf e}\} ) = 
\chi^{1}_{M_{3}} (\{ (p_{1},p_{2}),{\bf e}\} ) +
\xi^{1}_{M_{3} - 1} (\{ (p_{1},p_{2}),{\bf e}\} )
\end{equation}
where $\{ (p_{1},p_{2}),{\bf e}\} $ is a non -- oriented face of the
graph $G_{2}$ and the cochains $\chi^{1}_{i} = 0$ for 
$i < M^{\prime}_{3}(\chi^{1} )$ and $i > M_{3}(\chi^{1} )$.

Let the arbitrary cochains $\xi^{1}_{i} \in C^{1}(P(G_{2}),{\bf Z}^{add}_{2})$,
$i = M^{\prime}_{3},...,M_{3} - 1$, satisfying the equations (\ref{4.14})
be given. Since $\chi^{1} \in B_{1}(P(G_{3}),{\bf Z}^{add}_{2})
= Z_{1}(P(G_{3}),{\bf Z}^{add}_{2})$, the equations (\ref{4.14})
imply that the right hand sides of the equations (\ref{4.15}) -- (\ref{4.17})
are the cycles, i. e. they belong to the group 
$Z_{1}(P(G_{2}),{\bf Z}^{add}_{2})$. For the graph
$G_{2} = G(M^{\prime}_{1},M^{\prime}_{2};M_{1},M_{2})$ the group
$Z_{1}(P(G_{2}),{\bf Z}^{add}_{2})$ coincides with the group
$B_{1}(P(G_{2}),{\bf Z}^{add}_{2})$ and the group
$Z_{2}(P(G_{2}),{\bf Z}^{add}_{2})$ consists of the only cochain
which is equal to zero on any face. (The homology groups are trivial).
Therefore the cochains $\xi^{2}_{i} \in C^{2}(P(G_{2}),{\bf Z}^{add}_{2})$
are defined uniquely by the equations (\ref{4.15}) -- (\ref{4.17}).
Then the relations (\ref{3.2}), (\ref{3.5}) define uniquely the
cochain $\xi^{2} \in C^{2}(P(G_{3}),{\bf Z}^{add}_{2})$ satisfying the 
equation $\partial \xi^{2} = \chi^{1} $. Hence the cochain 
$\xi^{2} \in C^{2}(P(G_{3}),{\bf Z}^{add}_{2})$ satisfying the 
equation $\partial \xi^{2} = \chi^{1} $ is given uniquely by 
cochains $\xi^{1}_{i} \in C^{1}(P(G_{2}),{\bf Z}^{add}_{2})$,
$i = M^{\prime}_{3},...,M_{3} - 1$, satisfying the equations (\ref{4.14}).

The relations (\ref{3.2}), (\ref{3.5}) imply
\begin{eqnarray}
\label{4.18}
\left( \prod_{\{ p,e_{1},e_{2}\} \in P(G_{3})}
(\tanh \beta E(\{ {\bf p},{\bf e}_{1},{\bf e}_{2}\} ))^{\frac{1}{2} 
(1 - (- 1)^{\xi^{2} (\{ p,e_{1},e_{2}\} )})}
\right) |_{E(\{ p,(1,0,0),(0,1,0)\} ) \, = \, \infty } = \nonumber \\
\prod^{M_{3} - 1}_{i \, = \, M^{\prime}_{3}}
\prod_{\{ (p_{1},p_{2}),e\} \in P(G_{2})}
(\tanh \beta E(\{ (p_{1},p_{2},i),{\bf e},(0,0,1)\} ))^{\frac{1}{2} 
(1 - (- 1)^{\xi^{1}_{i} (\{ (p_{1},p_{2}),e\} )})}. 
\end{eqnarray} 
Now the relations (\ref{2.15}), (\ref{2.25}), (\ref{3.1}), (\ref{4.1}),
(\ref{4.18}) and Theorem 4.3 from the paper \cite{6} imply the 
relation (\ref{4.11}).

The relations (\ref{3.2}), (\ref{3.5}), (\ref{3.10}), (\ref{3.11})
imply
\begin{eqnarray}
\label{4.19}
\left( \prod_{\{ p,e_{1},e_{2}\} \in P(G_{3})}
(\tanh \beta E(\{ {\bf p},{\bf e}_{1},{\bf e}_{2}\} ))^{\frac{1}{2} 
(1 - (- 1)^{\xi^{2} (\{ p,e_{1},e_{2}\} )})}
\right) |_{E(\{ p,(1,0,0),(0,1,0)\} ) \, = \, - \, \infty } = \nonumber \\
(- 1)^{\sum^{M_{3}}_{i \, = \, M^{\prime}_{3}} 
\sum_{\{ (p_{1},p_{2}),e_{1},e_{2}\} \in P(G_{2})}
\xi^{2}_{i} (\{ (p_{1},p_{2}),e_{1},e_{2}\} )} \times \nonumber \\
\left( \prod^{M_{3} - 1}_{i \, = \, M^{\prime}_{3}}
\prod_{\{ (p_{1},p_{2}),e\} \in P(G_{2})}
(\tanh \beta E(\{ (p_{1},p_{2},i),{\bf e},(0,0,1)\} ))^{\frac{1}{2} 
(1 - (- 1)^{\xi^{1}_{i} (\{ (p_{1},p_{2}),e\} )})} \right). 
\end{eqnarray}
Summing up the relations (\ref{4.15}) -- (\ref{4.17}) we obtain
\begin{equation}
\label{4.20}
\partial \left( \sum^{M_{3}}_{i \, = \, M^{\prime}_{3}}
\xi^{2}_{i} \right) = \sum^{M_{3}}_{i \, = \, M^{\prime}_{3}} \chi^{1}_{i}.
\end{equation}
The group $Z_{2}(P(G_{2},{\bf Z}^{add}_{2})$ consists of the only cochain
which is equal to zero on any face. (The homology group is trivial).
Hence the equation (\ref{4.20}) defines the cochain
$\sum^{M_{3}}_{i \, = \, M^{\prime}_{3}} \xi^{2}_{i} $ uniquely. Now
the relations (\ref{2.15}), (\ref{2.25}), (\ref{3.1}), (\ref{4.1}),
(\ref{4.19}), (\ref{4.20}) and Theorem 4.3 from the paper \cite{6}
imply the relations (\ref{4.12}), (\ref{4.13}). The theorem is proved.

\end{document}